\documentclass[
amsmath,
amssymb,
aps, 
pra,
twocolumn, 
groupedaddress,
nobalancelastpage,
floatfix]
{revtex4-2}
\usepackage{mathtools}
\usepackage[colorlinks]{hyperref}
\usepackage{lipsum}
\usepackage{bbold}
\usepackage{bm}
\usepackage{dsfont}
\usepackage{enumerate}
\usepackage{graphicx}
\usepackage{float}
\usepackage {xcolor}
\usepackage{cleveref}
\usepackage{amsmath}
\usepackage{soul}
\usepackage[labelsep=period]{caption}
\usepackage{tabularx}
\newcolumntype{Y}{>{\centering\arraybackslash}X}
\usepackage[labelsep=period]{caption}
\newcommand{\nullarg}{{}\cdot{}}
\newcommand{\eqq}[1]{\begin{align} #1 \end{align}}

\DeclarePairedDelimiterX{\mean}[1]{\langle}{\rangle}{
	\ifblank{#1}{\nullarg}{#1}
}
\DeclarePairedDelimiterX{\abs}[1]{\lvert}{\rvert}{
	\ifblank{#1}{\nullarg}{#1}
}
\DeclarePairedDelimiterX{\norm}[1]{\lVert}{\rVert}{
	\ifblank{#1}{\nullarg}{#1}
}
\DeclarePairedDelimiterX{\bra}[1]{\langle}{\rvert}{#1}

\DeclarePairedDelimiterX{\ket}[1]{\lvert}{\rangle}{#1}
\DeclarePairedDelimiterX{\braket}[2]{\langle}{\rangle}{
	\ifblank{#1}{\nullarg}{#1} \delimsize\vert \ifblank{#2}{\nullarg}{#2}
}
\DeclarePairedDelimiterX{\sandwich}[3]{\langle}{\rangle}{
	\ifblank{#1}{\nullarg}{#1} \delimsize\vert \ifblank{#2}{\nullarg}{#2} \delimsize\vert \ifblank{#3}{\nullarg}{#3}
}
\DeclarePairedDelimiterX{\inner}[2]{\langle}{\rangle}{
	\ifblank{#1}{\nullarg}{#1} , \ifblank{#2}{\nullarg}{#2}
}

\begin{document}
	
    \title{Charge creation via quantum tunneling in one-dimensional Mott insulators: \\ A numerical study of the extended Hubbard model }
	
	\author{Thomas Hansen$^1$, Lars Bojer Madsen$^1$, and Yuta Murakami$^{2,3}$}
	\affiliation{$^1$Department of Physics and Astronomy, Aarhus University, Ny Munkegade 120, DK-8000 Aarhus C, Denmark\\
         $^2$ Institute for Materials Research, Tohoku University, Sendai 980-8577, Japan\\
          $^3$Center for Emergent Matter Science, RIKEN, Saitama 351-0198, Japan}
	\date{\today}
	\begin{abstract}
         Charge creation via quantum tunneling, i.e. dielectric breakdown, is one of the most fundamental and significant phenomena arising from strong light(field)-matter coupling.
In this work, we conduct a systematic numerical analysis of quantum tunneling in one-dimensional Mott insulators described by the extended ($U$-$V$) Hubbard model. We discuss the applicability of the analytical formula for doublon-holon (DH) pair production, previously derived for the one-dimensional Hubbard model, which highlights the relationship between the tunneling threshold, the charge gap, and the correlation length. We test the formulas ability to predict both DH pair production and energy increase rate.
Using tensor-network-based approaches, we demonstrate that the formula provides accurate predictions in the absence of excitonic states facilitated by the nearest-neighbor interaction $V$. However, when excitonic states emerge, the formula more accurately describes the rate of energy increase than the DH pair creation rate and in both cases gets improved by incorporating the exciton energy as the effective gap. 
	\end{abstract}
	
	\maketitle
	

	\section{Introduction} \label{Sec: Introduction}

    In strongly correlated systems (SCS), various intriguing quantum phenomena emerge from electronic correlations, as exemplified by high-$T_c$ superconductivity in cuprates \cite{TMO04, MMIT97, Tokura_RMP}.
Driving such systems out of equilibrium offers the potential for dramatic control over material properties and the observation of exotic optical responses that reflect many-body effects~\cite{Giannetti2016review, Basov2017review, Sentef2021nonthermal, Murakami2023_review}. In particular, with the advancement of strong-laser techniques in the THz and mid-infrared regimes, the study of strong-field physics in SCSs has become an increasingly important field.

    A particularly important and fundamental phenomenon is charge creation via quantum tunneling under a strong (DC) electric field~\cite{Kruchinin2018RMP}. Quantum tunneling can be regarded as the initial step in various intriguing processes. For instance, the charges created through tunneling can induce the metallization of a system, leading to a photo-induced insulator-to-metal transition~\cite{Yamakawa2017,Takamura2023}. The quantum tunneling is also closely related to high-harmonic generation (HHG), another fundamental phenomenon emerging from strong light-matter coupling. HHG has shown great potential both as a tool for producing ultrashort ($10^{-18}$ - $10^{-15}$ s) laser pulses \cite{Schubert2014, Lewenstein1994, Lein2003, Ghimire2011, Li2008} and as a spectroscopic technique \cite{Li2008, Lein2002, Torres2007, Kraus2015, Luu2018, Silva2018}. Previous theoretical studies have revealed that the basic mechanism of HHG in strongly correlated insulators can be understood using the three-step model for various types of charged elementary excitations emerging from strong correlations~\cite{Murakami2018, Murakami2021, Imai2020, Murakami2024HHG}. 
    This three-step model is conceptually similar to those developed for atomic gases~\cite{Corkum1993PRL} and semiconductors~\cite{Vampa2015PRB}, where the first step involves charge creation via quantum tunneling.
    Thus, the detailed understanding of the nature of quantum tunneling in SCSs is crucial for understanding a wide variety of subsequent or resultant non-equilibrium phenomena.

    Previously, to understand quantum tunneling effects in SCSs exposed to strong electric fields, the Hubbard model—one of the fundamental models for SCSs—has been extensively studied~\cite{Oka2003,Oka2005,Oka2010,Eckstein2010breakdown,Heidrich-Meisner2010,Oka2012,Lenarcic2012}. This model exhibits a Mott insulating phase at half-filling for sufficiently large local Coulomb interaction $U$, where all lattice sites are predominantly singly occupied. External perturbations can create doubly occupied sites (doublons) and unoccupied sites (holons), which are basic charge carriers of the system. 
    Numerical and analytical calculations have demonstrated that the doublon-holon (DH) production rate $\Gamma$ due to tunneling follows a threshold behavior:
    \eqq{
    \Gamma \propto F_0 \exp\left(-\frac{\pi F_{\rm th}}{F_0}\right), 
    }
    where $F_0$ is the strength of the applied electric field and $F_{\rm th}$ is the threshold field.
    In particular, by combining the Bethe ansatz solution with the Landau-Dykhne method, T. Oka analytically derived an expression for the proportionality factor and threshold field in the one-dimensional Hubbard model~\cite{Oka2012}.
    We refer to $F_{\rm th}$ calculated with this formula as $F_{\rm th}^{\rm Mott}$;  
    \begin{equation}
\begin{aligned}
       &\Gamma_{\rm Theory} =  \frac{F_0}{2\pi}\frac{|e| a}{\hbar} \exp\left(-\frac{\pi F_{\rm th}^{\rm Mott}}{F_0}\right), \\
    &F_{\rm th}^{\rm Mott} \simeq \frac{\Delta_{\rm Mott}}{2|e|\xi}, \label{eq: Gamma formula} 
\end{aligned}
\end{equation}
     where $\Delta_{\rm Mott}$ is the Mott gap, $e$ is the charge of an electron, $a$ is the systems bond length and $\xi$ is the DH correlation length. 
     A similar expression of the threshold field is also obtained for one-dimensional strongly correlated insulators using a low-energy effective model ~\cite{takasan2019}.
     Equation~(\ref{eq: Gamma formula}) provides an intuitive interpretation of the threshold field, making it a widely used tool in both experimental \cite{Yamakawa2017, Li2022, Mayer2015, Takamura2023} and theoretical \cite{Mayer2015, Tohoyama2023, AlShafey2023, AlShafey2024} studies for interpreting and analyzing results.
     For example, the correlation length of a DH pair has been estimated using Eq.~(\ref{eq: Gamma formula}) for a two-dimensional system~\cite{Yamakawa2017} as well as a ladder-type system~\cite{Tohoyama2023}. 
    However, strictly speaking, Eq.~(\ref{eq: Gamma formula}) is derived for the one-dimensional Hubbard model, which is integrable, 
    and its applicability to systems beyond this limit is not fully guaranteed. 
    
In this work, we explore charge creation processes due to quantum tunneling in SCSs beyond the standard Hubbard model and reveal the applicability of the above mentioned formula for DH creation. 
Namely, we focus on the one-dimensional extended ($U$-$V$) Hubbard model, for which the formula was not derived but which describe generic SCSs better than the standard Hubbard model. 
An important dynamical feature of this model is the possibility of a Mott exciton~\cite{Jeckelmann03}, which is a binding state of a DH pair.
The existence of such exciton states may change the excitation process of the Mott insulator, in particular, the tunneling process. We also discuss the ability of the above formula to capture the rate of
energy increase in the system.

	This paper is structured as follows. In Sec.~\ref{Sec: Theory}, we explain the theoretical background and numerical methods used to generate the result. The results are then presented and discussed in Sec.~\ref{Sec: Result}, which is finally followed by a summary, and outlook in Sec.~\ref{Sec: Summary}.

\section{Theory and methods} \label{Sec: Theory}
\begin{figure*}
\captionsetup{justification=raggedright}
    \centering
    \includegraphics[width=\linewidth]{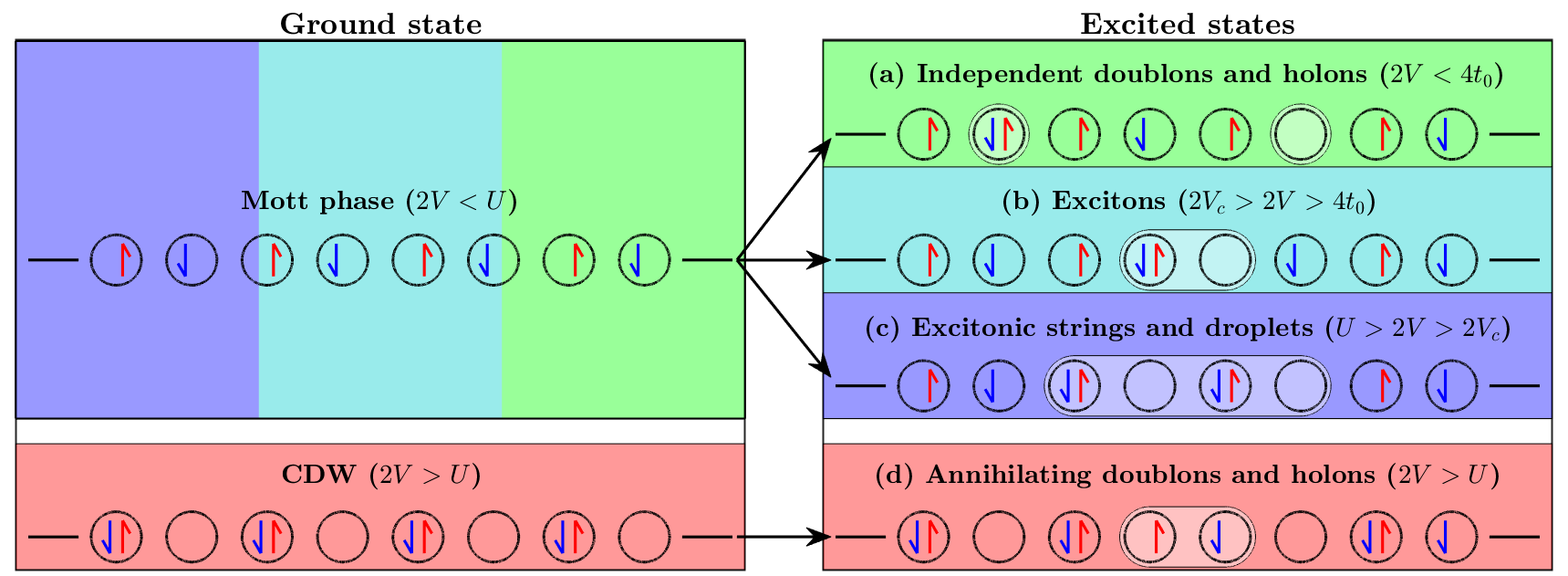}
     \caption{Schematic illustration of the regimes of the extended Hubbard model relevant to this work depending on the value of $V$ relative to $U$ ($U=10$). Each black ring represents a lattice site. An arrow in a circle indicates an electron, with a given spin occupying the corresponding lattice site. In the left panel the two phases of the ground state are illustrated by their dominant configurations. 
     In the right part, for excited states, each panel illustrates a different dynamical regime and a corresponding characteristic configuration for the excitation. In (b) and (c), $V_c$ denotes the critical $V$ value at which excitonic strings and droplets become significant to the system, see Sec.~\ref{Sec: Theory and Methods - The extended Hubbard model - 2V approx U but 2V<U}. 
     Lighter colors highlight the correlation among excited sites.
     The system treated here is infinitely long, which is represented by the short horizontal line at either end of the configurations.}
    \label{Fig: schema}
\end{figure*}
In this section, we cover the theoretical background and the numerical methods used in this work. We start by introducing the extended Hubbard model. Then, we discuss Eq.~(\ref{eq: Gamma formula}), which is followed by a description of the practical approaches we used to calculate the quantities needed to generate and analyze the results of this work. The section ends with an overview of the simulation values. Many of the points of this section are discussed in relation to the optical conductivities shown in the Appendix.

\subsection{The extended Hubbard model} \label{Sec: Theory and Methods - The extended Hubbard model}
We focus on the one-dimensional extended Hubbard model, which is a one-band model defined on a linear chain.
We describe this model using a natural unit system, in which $\hbar = |e| = a = t_0 = 1$, where $a$ is the bond length and $t_0$ is the hopping matrix element.
Within these constraints, the Hamiltonian can be written as \cite{Essler_Hubbard_book, Jeckelmann03}
\begin{align}
	\hat{H}(t) &=\hat{H}_\text{Hop}(t) + \hat{H}_U + \hat{H}_V\\
	\hat{H}_\text{Hop}(t) &= -\sum_{i,\sigma} \left(e^{iA(t)}\hat{c}_{i,\sigma}^\dagger \hat{c}_{i+1,\sigma}+h.c.\right) \label{eq: H_Hop}\\
	\hat{H}_U &=  U\sum_{i}\hat{n}_{i,\uparrow}\hat{n}_{i,\downarrow} \label{eq: H_U}\\
	\hat{H}_V &= V\sum_{i}\left(\hat{n}_{i,\uparrow}+\hat{n}_{i,\downarrow}\right)\left(\hat{n}_{i+1,\uparrow} +\hat{n}_{i+1,\downarrow}\right). \label{eq: H_V}
\end{align}
Here $\hat{H}_\text{Hop}(t)$ is the hopping term of the Hamiltonian, $\hat{H}_U$ is an on-site electron-electron interaction term, and $\hat{H}_V$ is a nearest-neighbor electron-electron interaction term. In Eq.~(\ref{eq: H_Hop}), $e^{iA(t)}$ is Peierls phase, with  $A(t)$ the vector potential of the external electric field applied along the chain direction. Note that the electric field $F(t)$ equals $-\partial_t A(t)$ and that both $A(t)$ and $F(t)$ are independent of position, in the electric dipole approximation. Finally for $\hat{H}_\text{Hop}(t)$, $\hat{c}_{i,\sigma}^\dagger$ ($\hat{c}_{i,\sigma}$) is the creation (annihilation) operator for an electron in a Wannier state centered on lattice site $i$ with spin $\sigma$. In Eq.~(\ref{eq: H_U}), $U$ is 
the value of the onsite Coulomb interaction,
and $\hat{n}_{i,\sigma}=\hat{c}_{i,\sigma}^\dagger\hat{c}_{i,\sigma}$ is the counting operator for electrons onsite $i$ and with spin $\sigma$. In Eq.~(\ref{eq: H_V}), $V$ is the value of the Coulomb interaction between nearest-neighboring sites.
Some cuprates such as Sr$_2$CuO$_3$~\cite{Imada1998} and organic compounds such as ET-F$_2$TCNQ~\cite{Hasegawa1997} can be described by the one-dimensional extended Hubbard model at half filling. 
Previous studies typically estimate \( U \) to be in the range \( 8 \lesssim U \lesssim 10\) for these materials~\cite{Sota2010,Ogasawara2000,Miyamoto2019}. 
Thus, in this paper, we fix $U=10$ as a typical value of actual materials and focus on the half-filled case at the temperature $T=0$.

Since the dynamical properties of the system depends on the ground state and its excitation structures, we first briefly summarize these points.
The ground states phase diagram of the 1D extended Hubbard model at half filling hosts several phases~\cite{Hirsch1984PRL,Jeckelmann2002PRL}. In particular, for $U\gg 1$ as in our case ($U=10$), it shows the Mott insulating phase ($U\gtrsim 2V$) and the charge density wave (CDW) phase ($U\lesssim 2V$), see the left-hand panel of Fig.~\ref{Fig: schema}. 
Intuitively speaking, the Mott insulating phase, in the upper left of Fig.~\ref{Fig: schema}, favors configurations containing only singly occupied sites due to the energy costs of making doubly occupied sites (doublons), which makes the system an insulator. 
It is characterized by quasi-long-range antiferromagnetic spin correlations due to spin-exchange coupling.
The CDW phase, seen in the lower left of Fig.~\ref{Fig: schema}, is distinguished by an alternating arrangement of doublons and holons, which saves the energy cost associated with the strong nearest-neighbor interaction $V$. Although the ground states for $U\gg 1$ can be categorized into these two phases, we can further classify the Mott insulating phase into four regimes depending on its excitation structures~\cite{Jeckelmann03}. These regimes are captured by the optical conductivity, and are summarized below and illustrated in the right-hand side of Fig.~\ref{Fig: schema}. 
In short, in the standard Hubbard model ($V=0$), the excitation structure is described by the quasi-particle picture of doulons and holons, while the nonlocal interaction $V$ yields the excitonic effects between them. For completeness, we show the optical conductivity obtained from our numerical analysis in the Appendix.

\subsubsection{Independent doublons and holons: $2V\lesssim 4$} \label{Sec: Theory and Methods - The extended Hubbard model - 2V<4t0}
In this regime, shown in Fig.~\ref{Fig: schema} (a), as in the case of \( V = 0 \), the relevant excited states are primarily described within the quasi-particle picture of independent doublons and holons. 
Reflecting this situation, the doublon and holon are separated in Fig.~\ref{Fig: schema} (a). This independence means the doublons and holons are free to move about the lattice creating a DH continuum. 
The energy gap between the ground state and the bottom of that continuum is the Mott gap \( \Delta_{\rm Mott} \).
The non-local Coulomb interaction \( \hat{H}_V \) of Eq.~\eqref{eq: H_V} induces an attractive interaction between a doublon and a holon. 
However, for \( 2V \lesssim 4 \), this attraction is not strong enough to overcome the loss of kinetic energy from \( \hat{H}_{\rm Hop}(t) \) of Eq.~\eqref{eq: H_Hop} which would result from binding the doublon to the holon. 
As a result, no bound state is formed, and the doublon and holon remain effectively independent.
The cases of \( V = \{0, 1, 2\} \) for \( U = 10 \), which we analyze later, correspond to this regime.

\subsubsection{Mott exciton: $U > 2V_c > 2V \gtrsim 4$} \label{Sec: Theory and Methods - The extended Hubbard model - U>2V>4}
In this regime the attractive interaction between a doublon and a holon is strong enough to form a bound state, known as a Mott exciton, as illustrated in Fig.~\ref{Fig: schema} (b). 
Here, \( V_c \) (\( > \mathcal{O}(t_0=1) \)) is a critical value determined by the energy cost of creating an additional DH pair on top of a single Mott exciton~\cite{Jeckelmann03}. 
The signature of the Mott exciton in the linear optical spectrum appears as a peak at \( \omega = \Delta_{\rm exc} \) below the Mott gap \( \Delta_{\rm Mott} \) (see Appendix). 
Note that even in this regime, a continuum originating from unbound DH pairs exists above the Mott gap. 
In practice, the cases of \( V = \{3, 4\} \) for \( U = 10 \), which we analyze later, correspond to this regime.

\subsubsection{Exciton string and CDW droplet: $U\gtrsim 2V >2V_c$} \label{Sec: Theory and Methods - The extended Hubbard model - 2V approx U but 2V<U}

With a further increase in \( V \), collective excited states emerge below the Mott gap, which can be regarded as compositions of multiple Mott excitons.
One such type is the so-called exciton string, which consist of \( n_{\rm exc} \) Mott excitons bound together (\( n_{\rm exc} \)-exciton string), as illustrated in Fig.~\ref{Fig: schema} (c). 
These excitations appear in the regime not too close to \( U = 2V \). In the optical spectrum, several peaks below the Mott gap correspond to \( n_{\rm exc} \)-exciton strings with \( n_{\rm exc} = 1, 2, 3, \dots \).
Another type is the so-called CDW droplets, which can be regarded as a superposition of exciton strings with different sizes.
These excitations appear around \( U = 2V \), as a means to reduce kinetic energy through size fluctuations of exciton strings. 
CDW droplets exhibit a continuous band corresponding to the range of size fluctuations, which is reflected as a broad spectral weight below the Mott gap in the optical spectrum.
In practice, for \( U = 10 \), only the CDW droplets are relevant, and the case of \( V = 5 \) considered below corresponds to this regime.

 \subsubsection{Charge density wave phase: $2V\gtrsim U$} \label{Sec: Theory and Methods - The extended Hubbard model - 2V>U}
For completeness, we also comment on the excitations from the CDW phase, which favors an alternating configuration of doublons and holons.
Here, these excitations are described as the annihilation of a doublon and a holon, as illustrated in Fig.~\ref{Fig: schema} (d). 
The illustration shows that the spin-up and spin-down electrons from the annihilated doublon cannot move freely due to the large energy cost associated with \( V \). 
This implies that no continuum exists for the excitations in the CDW phase, which we have numerically confirmed in the optical spectrum (not shown).

Due to the fundamental differences between the Mott phase and the CDW phase, the tunneling formula in Eq.~\eqref{eq: Gamma formula} is not expected to be applicable. We have conducted tests of this similar to those presented in Sec.~\ref{Sec: Result}, and indeed, they confirm the expected lack of applicability, in spite of $\Gamma$ still showing threshold-like behavior. 
In the following, we focus on the Mott phase and numerically analyze the applicability of the formula across different dynamical regimes.


\subsection{Methods} \label{Sec: Theory - methods and systems}
We use two tensor-network-based methods to handle the extended Hubbard model. The first one is the infinite time-evolving block decimation (iTEBD) \cite{Kjall13, Vidal07} and the second is the density matrix renormalization group (DMRG)\cite{ITensor_main}. 
Both methods are based on matrix product states (MPS). All results except the $\Delta_{\rm Mott}$ values are obtained using iTEBD. 
The iTEBD directly deals with systems in the thermodynamic limit using the translational invariance of MPS.
In iTEBD, we first make imaginary time propagation to find the ground state, for which we use a cutoff dimension of 1000, and then we operate the real-time simulation. 
The imaginary time propagation is done with decreasing time step size, ending with $20,000$ steps at $d\tau = 0.005$. The following real-time propagation use a cutoff dimension of at least 1500 with a time step size of $0.02$. 
We have thoroughly tested that this combination of step count and step size is sufficient to obtain convergence.
In the iTEBD simulations, we use the conservation of spin and electron count to speed up the simulations. Furthermore, we apply a staggered magnetic field small enough to keep relevant physical observables unchanged but to open a small spin gap. The latter makes the imaginary time propagation efficient.

DMRG employs the variational principle to find the ground state. 
We use it to obtain the ground-state energy of finite-size systems in order to evaluate \( \Delta_{\rm Mott} \), as discussed below in Sec.~\ref{sec:Mott_gap}. 
In practice, we set the cutoff dimension of the MPS to 1000 and the system size to 160 sites, where we observe good convergence of \( \Delta_{\rm Mott} \) with respect to both the cutoff dimension and the system size.

\subsection{The analytical rate formula} \label{Sec: Theory and methods - The Gamma formula}

The formula for the DH production rate, Eq.~(\ref{eq: Gamma formula}), was derived in two steps \cite{Oka2012}.  
First, Landau-Dykhne theory~\cite{Dykhne1962,Davis1976} was combined with the Bethe ansatz solution of the non-Hermitian Hubbard model~\cite{Fukui1998} to evaluate the tunneling probability from the ground state to  states with a doublon-holon pair. This yields the exponential part of Eq.~(\ref{eq: Gamma formula}), namely, the $\exp\left(-\frac{\pi F_{\rm th}^{\rm Mott}}{F_0}\right)$ term.  
Secondly, physical considerations regarding the repetition rate of tunneling events and the relation between the created charge and a doublon were used to determine the prefactor of the formula.
Note that from its derivation, we can expect that the same form of Eq.~\eqref{eq: Gamma formula} can also describe the evolution of different physical observables other than the doublon count, although the constant factor may be largely different. For example, as a doublon-holon pair is created, the energy of the system increases accordingly. Thus, it is reasonable to expect that Eq.~(\ref{eq: Gamma formula}) could capture the change in energy.

Still, we have to keep in mind that the derivation of Eq.~(\ref{eq: Gamma formula}) comes  with certain limitations.
Firstly, Landau-Dykhne theory can be viewed as a leading-order asymptotic approximation that describes the transition rate to exponential accuracy, but neglects the prefactor and correction terms. Identification of the prefactor and correction terms can be done by asymptotic methods \cite{Olver} as recently considered in the context of tunneling ionization of atoms and molecules in the weak-field asymptotic theory \cite{Tolstikhin2011,Trinh2013,Tolstikhin2014}.
Secondly, in order to translate the excitation rate to doublon creation rate, it was assumed that the difference in doublon count between the ground state and excited state was exactly one. Thirdly, the production of multiple pairs and the possible annihilation processes of preexisting DH pairs are ignored \cite{Oka2012}. The first point limits the applicability of the formula's general precision, presumably mainly at higher field strengths, where the prefactor becomes increasingly important.  The second point effectively limits the formula to $U\gg \{V, t_0=1\}$, i.e., to Mott insulators, see Sec.~\ref{Sec: Theory and Methods - The extended Hubbard model - 2V<4t0}. 
The third point limits the degree to which one can excite the system. 
Namely, if the system is strongly excited to have several DH pairs, they affect the subsequent tunneling processes, which is not considered in the formula.
This mainly limits the field strengths that can be used, but also limits the duration over which Eq.~\eqref{eq: Gamma formula} remains accurate for any given field strength. 

As stated in the introduction we wish to test the validity of this formula for $V > 0$, i.e., outside the realm in which it was derived. To test this we will need to generate $\Gamma$ values via simulations, $\Gamma_{\rm Sim}$, and compare with the right-hand side of Eq.~(\ref{eq: Gamma formula}), denoted $\Gamma_{\rm Theory}$. To calculate $\Gamma_{\rm Theory}$ values of $\Delta_{\rm Mott}$ and $\xi$ are needed. Below we will outline how we obtain these values. 
\subsubsection{Determining $\Gamma_{\rm Sim}$} \label{sec:get_gamma_sim}
$\Gamma_{\rm Sim}$ can be calculated by simulating the system under the effect of a DC field, specifically by studying the doublon count ($N_{\rm D}$). $N_{\rm D}$ is defined as $N_{\rm D} = \langle \hat{n}_{i,\uparrow}\hat{n}_{i,\downarrow}\rangle$. 
Note that which site $i$ is used is irrelevant as all sites are identical.
Furthermore, at the half-filling condition, the number of doublons is equal to that of holons.
Thus, $N_{\rm D}$ practically counts the number of DH pairs.
Assuming Eq.~(\ref{eq: Gamma formula}) is correct, then the rate at which the doublon count rises should be constant when the system is subjected to a DC field. This implies that the DH pair count ($N_{\rm D}$) satisfies $N_{\rm D}\propto 1-\exp(-\Gamma t)$, and should rise linearly for small $\Gamma t$. The inclination of that initial linear rise is $\Gamma_{\rm Sim}$, which we determine as the inclination of a straight line fitted to the initial linear rise. 
Similar to the procedure of Ref.~\cite{Eckstein2010breakdown}, the DC field is applied via the following form,
\begin{align}
A(t) = \begin{cases}
0 & t<0\\
F_0\frac{t}{2} - \frac{9F_0 t_r}{16\pi}\sin\left(\frac{\pi t}{t_\text{r}}\right) + \frac{F_0t_r}{48\pi} \sin\left( \frac{3\pi t}{t_\text{r}}\right) & 0\leq t < t_\text{r} \\
F_0 \left(t - \frac{t_r}{2} \right) & t\geq t_{\rm r} 
\end{cases}
\end{align}
Here $t_\text{r}$ is the ramp time of the field.
This field is a DC field for $t\geq t_{\rm r}$. If the ramp had been removed $F(t)$ would be discontinuous resulting in quench effects. Quench effects cause the doublon count to oscillate while rising, which complicates the fitting procedure used to extract $\Gamma_{\rm Sim}$ markedly. However, if $t_{\rm r}$ becomes too large the system will be driven out of the ground state by the time $t=t_{\rm r}$. So an appropriate $t_{\rm r}$ needs to be found. The second issue is in choosing $F_0$. Too small an $F_0$ means essentially no DH pair creation which makes numerically determining the actual DH pair creation difficult. Too large an $F_0$ causes the formula to break down by increasing the excitation during the field ramping. So an appropriate range of $F_0$'s also needs to be identified. We will demonstrate how we determined this value and range in Sec.~\ref{Sec: Result}(a).

\subsubsection{Determining $\Delta_{\rm Mott}$} \label{sec:Mott_gap}
The Mott gap $\Delta_\text{Mott}$ corresponds to the band gap between the upper and lower Hubbard bands in the single-particle spectrum. Numerically, we evaluate $\Delta_\text{Mott}$ as 
\begin{align}
	\Delta_\text{Mott} &= E_\text{GS}^{L+1}(U,V) + E_\text{GS}^{L-1}(U,V) - 2E_\text{GS}^{L}(U,V), \label{eq: Mott gap}
\end{align}
where $L$ is the number of the sites and $E_\text{GS}^{N_e}(U,V)$ is 
the ground state energy of the system with $N_e$ electrons 
with a minimal spin imbalance.
Note that $N_e=L$ corresponds to half-filling.
If we focus on the Mott insulating regime at $2V\lesssim U$, the above expression can be thought of as the summation of the minimum energy to create one independent doublon ($E_\text{GS}^{L+1}(U,V)-E_\text{GS}^{L}(U,V)$) and that to create one independent holon ($E_\text{GS}^{L-1}(U,V)-E_\text{GS}^{L}(U,V)$). Thus, it sets the bottom of the DH continuum. However, this quantity does not take into account the interaction between a doublon and a holon, and hence misses the excited states associated with it, like excitons, exciton strings and CDW droplets~\cite{Jeckelmann03}.

\subsubsection{Determining $\xi$}

The correlation length $\xi$ corresponds to the decay of a correlation function $G({|i-j|})= \langle \hat{c}^\dagger_{i\sigma} \hat{c}_{j\sigma} \rangle$ as 
\begin{align}
G(|l|) \sim \exp(-|l|/\xi).
\end{align}
in the large $|l|$ limit.
We numerically evaluate this correlation length $\xi$ with iTEBD as explained in Ref.~\cite{Kjall13}.
Namely, the correlation length can be associated with an eigenvalue of the transfer matrix which can be calculated from the matrix product states used in iTEBD.

\subsubsection{Resultant $F_{\rm th}^{\rm Mott}$ values}
Now that methods for calculating both $\Delta_{\rm Mott}$ and $\xi$ have been established, we can provide values for $F_{\rm th}^{\rm Mott}$. In Tab.~\ref{Tab: F_th Mott} those values are given for integer values of $V$ between 0 and 5, where we also show the obtained values for $\Delta_{\rm Mott}$ and $\xi$.

\begin{table}[htbp]
	\captionsetup{justification=raggedright}
\begin{tabularx}{\linewidth}{YYYY}
	\hline\hline
	$V$ & $F_\text{th}^{\rm Mott}$ &      $\Delta_{\rm Mott}$ &      $\xi$  \\
	\hline
	0  &  3.073 & 6.553 &  1.066 \\
	
    1  &  2.913 & 6.542 &  1.123 \\
    
    2  &  2.695 & 6.472 &  1.201 \\
    
    3  &  2.370 & 6.271 &  1.323 \\
    
    4  &  1.842 & 5.735 &  1.557 \\
    
    5  &  0.678 & 3.561 &  2.626 \\
    \hline\hline
\end{tabularx}
	\caption{$F_\text{th}^{\rm Mott}$, $\Delta_{\rm Mott}$, and $\xi$ values for the relevant values of $V$ in the Mott phase with $\Delta_\text{Mott}$ calculated by DMRG and Eq.~(\ref{eq: Mott gap}). $V$ and $\Delta_{\rm Mott}$ is given in units of $t_0$, $F_{\rm th}^{\rm Mott}$ is in units of $\frac{t_0}{a|e|}$, and $\xi$ is in units of $a$.}
\label{Tab: F_th Mott}
\end{table}

\subsection{Effects of excitons on $F_{\rm th}$} \label{Sec: Theory and methods - Alternative energy gap}
As described in Sec.~\ref{Sec: Theory and Methods - The extended Hubbard model} when $V_c>V>2$ an Mott exciton exists within the band gap of the Hubbard model.
Intuitively $\Delta_{\rm Mott}$, as it is used in Eq.~(\ref{eq: Gamma formula}), describes the energy required to generate a doublon. Since the excitonic states also include a doublon and have lower energy than $\Delta_{\rm Mott}$, it may be that the exciton energy, $\Delta_{\rm exc}$, better represent the energy required to produce a doublon than $\Delta_{\rm Mott}$ in this regime. For that reason, we calculate $\Delta_{\rm exc}$ as an ad-hoc alternative to $\Delta_{\rm Mott}$ in those cases, see Appendix. Using the $\Delta_{\rm exc}$ values in place of $\Delta_{\rm Mott}$ in the formula for $F_{\rm th}^{\rm Mott}$ provides a set of $F^{\rm exc}_{\rm th}$ values for the $V_c>V>2$ cases. 
Around $U\simeq 2V$, the Mott excitons are no longer well defined as CDW droplets emerge. In this regime, the first peak in the optical spectrum is called $\Delta_{\rm exc}$, although it loses its meaning as the exciton energy. We also generate associated $F_{\rm th}^{\rm exc}$ values for the $U\simeq 2V$ systems by utilizing these $\Delta_{\rm exc}$ values.
The obtained values of $\Delta_{\rm exc}$ and $F_{\rm th}^{\rm exc}$ can be found in Tab.~\ref{Tab: F_th exciton} for the integer $V$ values between $3$ and $U/2 = 5$.

\begin{table}[htbp]
	\captionsetup{justification=raggedright}
\begin{tabularx}{\linewidth}{YYY}
	\hline\hline
	$V$ & $F_\text{th}^{\rm exc}$ &      $\Delta_{\rm exc}$  \\
	\hline
	3  &  2.290 &  6.021 \\
	
    4  &  1.576 &  4.870 \\
    
    5  &  0.409 &  2.109 \\
    \hline\hline
\end{tabularx}
	\caption{$F_\text{th}^{\rm exc}$ and $\Delta_{\rm exc}$ for the relevant values of $V$ in the Mott phase. $\Delta_{\rm exc}$ is evaluated from the optical spectrum using iTEBD. $V$ and $\Delta_{\rm exc}$ are both in units of $t_0$ and $F_{\rm th}^{\rm exc}$ is in units of $\frac{t_0}{a|e|}$.}
\label{Tab: F_th exciton}
\end{table}

\section{Results}\label{Sec: Result}
In this section, we present and analyze the results of this work. 
In Sec.~\ref{sec:result_A}, focusing on the standard Hubbard model ($V=0$), we determine a value of $t_r$ and range of $F_0$ to compare numerical results and the analytical formula. 
In Sec.~\ref{sec:result_B}, we test the applicability of Eq.~(\ref{eq: Gamma formula}) on the extended Hubbard model in the Mott phase following the increase ratio of the doublon number. 
In Sec.~\ref{sec:result_C}, we address the applicability of the functional form of Eq.~\eqref{eq: Gamma formula} to describe the rate of change in the energy of the system.

\subsection{Lesson from the standard Hubbard model ($V=0$)} \label{sec:result_A}
In this section, we focus on the case of the standard Hubbard model ($V=0$) and identify a value for $t_r$ and range of $F_0$ values for comparing numerical results with the analytical formula~\eqref{eq: Gamma formula}. As pointed out in Sec.~\ref{sec:get_gamma_sim}, $t_{\rm r}$  needs to be neither so short that quench effects become substantial nor so large that the system leaves the ground state during the ramp. Furthermore, $F_0$ needs to be small enough to ensure the formula remains accurate and the system is not driven out of the ground state during the ramp but not so small that $\Gamma_{\rm Sim}$ becomes difficult to determine accurately. 
Using the simulation outlined in Sec.~\ref{Sec: Theory and methods - The Gamma formula} 
to determine $\Gamma_{\rm Sim}$ for $V=0$, where Eq.~\eqref{eq: Gamma formula} should be reliable, we identify suitable ranges of $t_r$ and $F_0$.

\begin{figure}
	\captionsetup{justification=raggedright}
	\includegraphics[width=\linewidth]{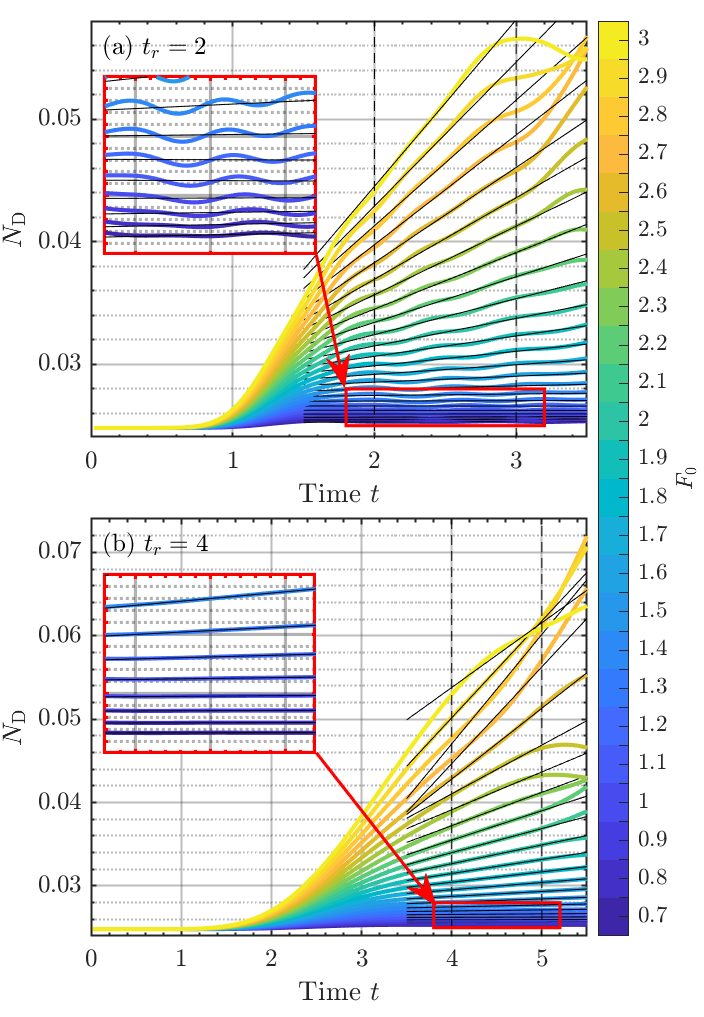}
	\caption{Doublon counts for the $U=10$ and $V=0$ system and (a) $t_r=2$ and (b) $t_r=4$, respectively. The colored lines indicate the value of the field strength, as indicated on the color bar. In this case $F_\text{th}^{\rm Mott} = 3.07$. The thin black lines are the linear fits to the doublon counts. We fit to the doublon counts between the vertical black dashed lines, i.e., after $t_\text{r}$ and for 1 unit of time. Inserts highlight the sensitivity of $N_D$ to $t_r$. The unit of time used here is $\frac{\hbar}{t_0}$ and the unit of $F_0$ is $\frac{t_0}{a|e|}$.}
	\label{Fig1: DH rates}
\end{figure}

Figure~\ref{Fig1: DH rates} depicts doublon counts of the $U=10$ and $V=0$ system for various field strengths at $t_r = 2$ and $t_r = 4$ in panel (a) and (b), respectively. The former choice of $t_{\rm r}$ is closer to the sudden quench protocol. The inserts in each panel show zooms of the region in the red boxes. Due to the wide range of frequency components in the electric field when $t_{\rm r} = 2$, the doublon counts show oscillatory behavior. This behavior makes it difficult to determine the increase ratio of the doublon counts by fitting. For a longer ramp time, the oscillatory behavior becomes less severe as expected, see the insert of Fig.~\ref{Fig1: DH rates} (b). 
Still, if the field strength is too low, the doublon count essentially remains constant
with small oscillatory behavior on top. 
This makes it difficult to accurately determine the associated $\Gamma_{\rm Sim}$. 
On the other hand, if the field strength becomes too large we start to see the doublon count bending during the fitting region. This non-linearity appears because of the state being driven out of the field-free ground state resulting in a different response to the field and therefore a different DH generation rate. Note that it cannot just be because the ground state gets depleted as time progresses, as that could only cause $N_D$ to decelerate, and acceleration is also observed. 

 \begin{figure}
 	\captionsetup{justification=raggedright}
 	\includegraphics[width=\linewidth]{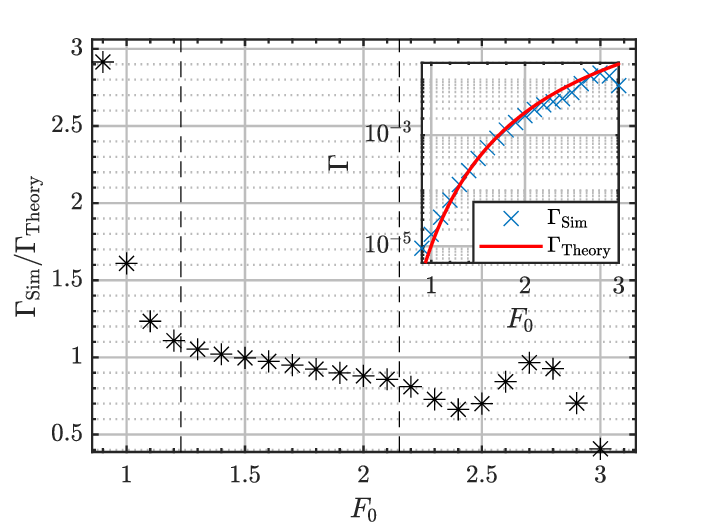}
 	\caption{Comparison between  $\Gamma_\text{Sim}$ and $\Gamma_\text{Theory}$ for $U=10, V=0$ and $t_r = 4$. The insert shows the $\Gamma_{\rm Sim}$ values obtained as blue crosses, overlaid by Eq.~\eqref{eq: Gamma formula} in red. On the main plot, we see the ratios between $\Gamma_{\rm Sim}$ and $\Gamma_{\rm Theory}$. When the ratio is 1, the simulation and formula agree perfectly. The vertical dashed lines indicate $F_0=0.4F_{\rm th}^{\rm Mott}$ and $F_0=0.7F_{\rm th}^{\rm Mott}$. The unit of $F_0$ used here is $\frac{t_0}{a |e|}$, and the unit of $\Gamma$ is $\frac{t_0}{\hbar}$.}
 	\label{Fig2: Gamma V0 comp}
 \end{figure}

In Fig.~\ref{Fig2: Gamma V0 comp} we show the ratio between $\Gamma_\text{Sim}$ and $\Gamma_\text{Theory}$ for $t_{\rm r}=4$. In the insert of Fig.~\ref{Fig2: Gamma V0 comp} we show a direct comparison between $ \Gamma_\text{Sim}$ and $\Gamma_\text{Theory}$, which are marked with blue crosses and a red line respectively. 
$\Gamma_{\rm Sim}$ and $\Gamma_{\rm Theory}$ agree reasonably well except for the lowest and highest field strengths.
At the lowest field strengths around $F_0 \lesssim 1.2$, due to the uncertainty of the fitting procedure mentioned above and/or that of the numerical simulation, the obtained $\Gamma_{\rm Sim}$ largely overestimates the expected value of $\Gamma_{\rm Theory}$.
In the high-$F_0$ regime around $F_0\gtrsim 2.2$, the deviation should result from excitation away from the ground state not captured by the formula. This deviation also suggests the reasonable agreement between simulation and Eq.~\eqref{eq: Gamma formula} around $F_0=2.7$ is mere coincidence. For $1.2 \lesssim F_0 \lesssim 2.2$, Eq.~(\ref{eq: Gamma formula}) appears to work well. However, a slight downwards tilt is observed in the ratios. This tilt is mainly a combination of the uncertainty in the fitting procedure resulting in an overestimation, which dominates at lower field strengths. At higher field strengths, the state gets driven out of the ground state resulting in the underestimation of $\Gamma_{\rm Sim}$. This uncertainty at lower field strengths gets amplified by the rapid drop in $\Gamma_{\rm Theory}$, as seen in the insert of Fig.~\ref{Fig2: Gamma V0 comp}, and by quench effects. As can be seen in the insert of Fig.~\ref{Fig1: DH rates} (a) quench effects result in oscillatory behavior in $N_{\rm D}$. This oscillatory behavior complicates the fitting procedure. In our results, the magnitude of those oscillations drops significantly slower than the overall incline, resulting in an increasing relative impact on the $\Gamma_{\rm Sim}$. Another final thing that could impact the ratios is the prefactor of Eq.~\eqref{eq: Gamma formula}. That prefactor was obtained via physical arguments.

To summarize, the results of Fig.~\ref{Fig2: Gamma V0 comp} suggest that Eq.~(\ref{eq: Gamma formula}) works reasonably well 
in the range of $1.2\lesssim  F_0\lesssim 2.2$, i.e. $0.4\lesssim F_0/F_\text{th}\lesssim 0.7$, given the limitations of both the derivation and the comparison method. 
In the following analysis, we use this range of field  rescaled by the expected threshold field $F_\text{th}$, i.e. $0.4\lesssim F_0/F_\text{th}\lesssim 0.7$, as a primary range where we compare numerical results with the analytical formula~\eqref{eq: Gamma formula} for different system parameters.  Here $F_{\rm th}$ will be either $F_{\rm th}^{\rm Mott}$ (Tab.~\ref{Tab: F_th Mott}) or $F_{\rm th}^{\rm exc}$ (Tab.~\ref{Tab: F_th exciton}) depending on which case we are addressing.

 \begin{figure}
 	\captionsetup{justification=raggedright}
 	\includegraphics[width=\linewidth]{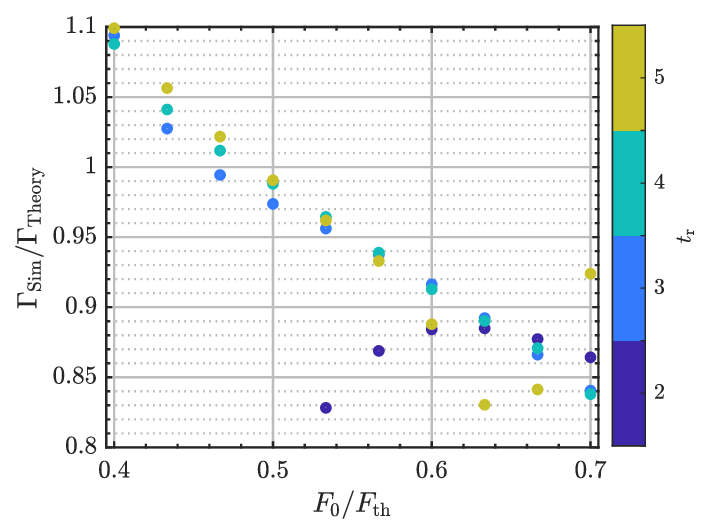}
 	\caption{Ratios of $\Gamma_\text{Sim}$ to $\Gamma_\text{Theory}$ for the $U=10$ and $V=0$ system with field strengths picked in accordance with Fig.~\ref{Fig2: Gamma V0 comp}. The ratios are shown for four different values of $t_r$, as indicated by the color bar. The unit of time used here is $\frac{\hbar}{t_0}$.}
 	\label{Fig: DH rates t_r}
 \end{figure} 

\begin{figure*}
    \captionsetup{justification=raggedright}
 	\includegraphics[width=\linewidth]{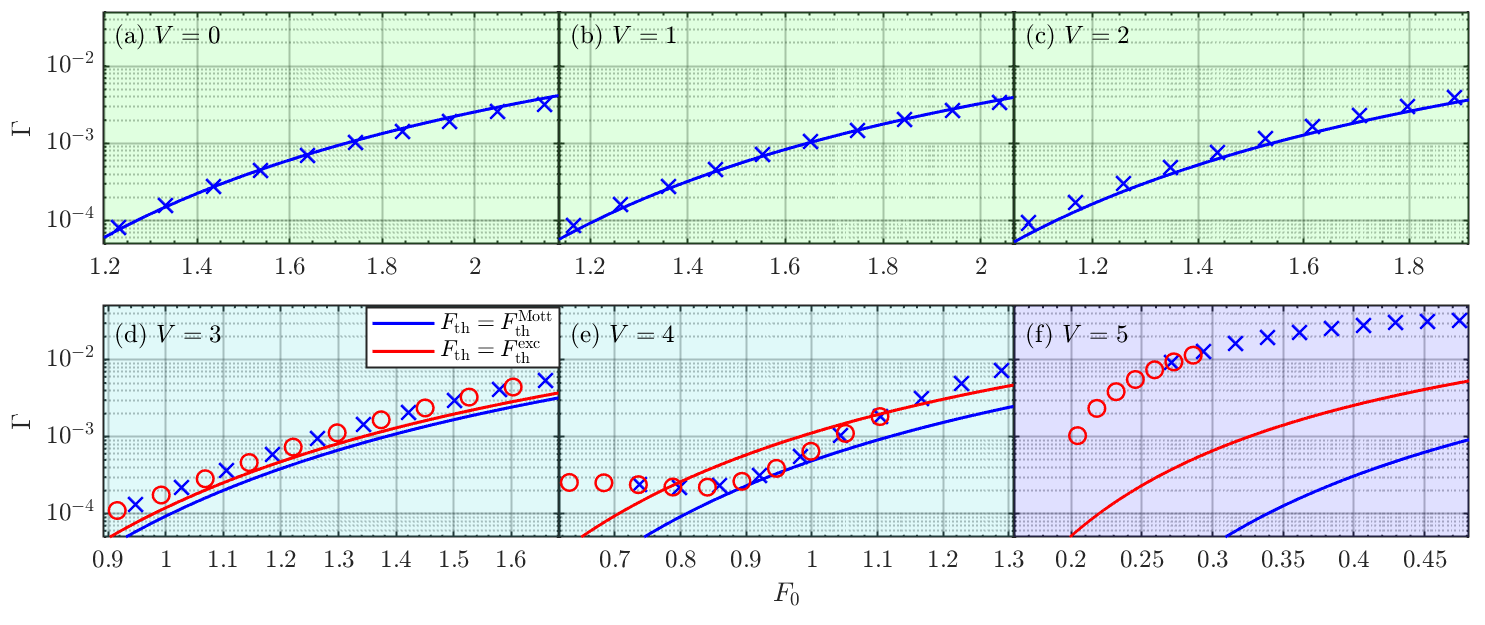}
 	\caption{Comparisons between $\Gamma_{\rm Sim}$ and $\Gamma_{\rm Theory}$ for integer values of $V$ in the Mott phase at $U=10$.
    The $\Gamma_{\rm Sim}$ values are evaluated in the range of $0.4F_{\rm th}\leq F_0 \leq 0.7F_{\rm th}$, where we use either $F_{\rm th} = F_{\rm th}^{\rm Mott}$ (blue crosses), or $F_{\rm th} = F_{\rm th}^{\rm exc}$ (red circles). 
    $\Gamma_{\rm Theory}$ are given by the fully drawn lines where the blue and red colors correspond to  $F_{\rm th} = F_{\rm th}^{\rm Mott}$ and  $F_{\rm th} = F_{\rm th}^{\rm exc}$ respectively. 
       The background colors indicate the regime each panel belongs to, see Fig.~\ref{Fig: schema}. $F_0$ is given in units of $\frac{t_0}{a|e|}$, and $\Gamma$ is in units of $\frac{t_0}{\hbar}$.}
 	\label{Fig V scan Gamma dir comp}
\end{figure*}
To demonstrate the impact of the choice of $t_{\rm r}$, we show ratios between $\Gamma_\text{Sim}$ and $\Gamma_\text{Theory}$ for $0.4\leq F_0/F_\text{th}\leq 0.7$ and $2\leq t_r \leq 5$ in Fig.~\ref{Fig: DH rates t_r}. We see that the ratios for $t_r = 2$ are consistently lower than 1, particularly for the smallest field strengths.
The deviation originates from the quench effects as mentioned above. Both the $t_r=3$ and $t_r=4$ results show a consistent almost linear tilt for all field strengths showed. The $t_r=5$ results shows that same tilt for $F_0/F_{\rm th}<0.6$ but deviates from that afterwards, resultng in worse agreement. As a result it appears that both $t_r=3$ and $t_r=4$ would be reasonable choices. We have chosen to continue with $t_r=4$.

We wish to note that the validity of the Eq.~\eqref{eq: Gamma formula} has been checked numerically for finite-size systems with a sudden quench in the standard Hubbard model~\cite{Oka2012}. Our systematic numerical analysis for the system in the thermodynamic limit further confirms the equations' validity and provides a revised framework for discussing its validity compared to the sudden quench.

\subsection{Comparing $\Gamma_{\rm Theory}$ and $\Gamma_{\rm Sim}$} \label{sec:result_B}

\begin{figure}
    \captionsetup{justification=raggedright}
    \includegraphics[width=\linewidth]{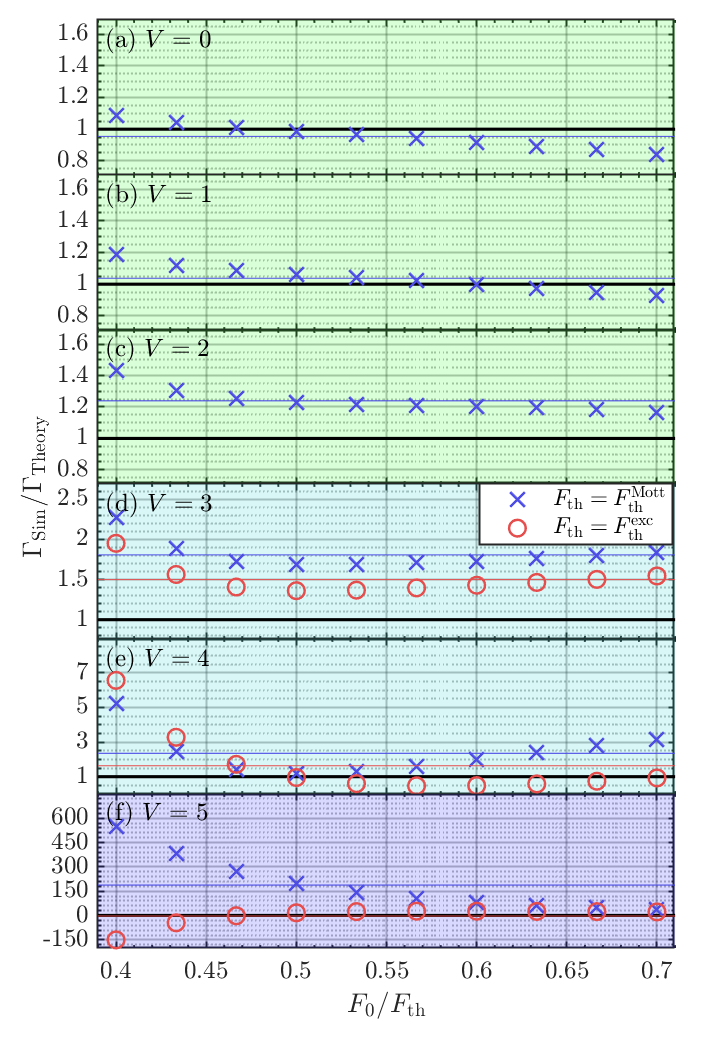}
    \caption{Ratios between $\Gamma_{\rm Sim}$ and $\Gamma_{\rm Theory}$ for integer values of $V$ in the Mott phase. The background color indicates the dynamical regime in question, see Fig.~\ref{Fig: schema}. Two sets of results are included, the blue crosses for $F_{\rm th}=F_{\rm th}^{\rm Mott}$ and the red circles for $F_{\rm th}=F_{\rm th}^{\rm exc}$. As the exciton is only separated from the DH continuum for $V>2$ we only include the $F_{\rm th}=F_{\rm th}^{\rm exc}$ results in panel (d-f). The averages of the ratios are indicated with thin blue and red lines respectively, and the ratio of 1 is indicated with a black horizontal line. Note the changing ordinates in panels (d-f).}
    \label{fig: Gamma comp ratios}
\end{figure}

Now that $t_r$ and $F_0/F_{\rm th}$ have been chosen, we move on to comparing $\Gamma_{\rm Sim}$ and $\Gamma_{\rm Theory}$ also for the $V\neq 0$ cases. 
Remember that $\Gamma_{\rm Sim}$ is extracted from the evolution of doublon counts $N_{\rm D}$, and 
that $\Gamma_{\rm Theory}$ is the threshold value evaluated from Eq.~\eqref{eq: Gamma formula} using either $F_{\rm th} = F_{\rm th}^{\rm Mott}$ or $F_{\rm th} = F_{\rm th}^{\rm exc}$.
We start with Figs.~\ref{Fig V scan Gamma dir comp} and \ref{fig: Gamma comp ratios}. In Fig.~\ref{Fig V scan Gamma dir comp}, direct comparisons between $\Gamma_{\rm Sim}$ and $\Gamma_{\rm Theory}$ are shown whereas in Fig.~\ref{fig: Gamma comp ratios} we show the ratio $\Gamma_{\rm Sim}/\Gamma_{\rm Theory}$, in both cases for all integer values for $0\leq V \leq 5$. In Figs.~\ref{Fig V scan Gamma dir comp} (a) and \ref{fig: Gamma comp ratios} (a), both $V=0$, we see good agreement between $\Gamma_{\rm Sim}$ and $\Gamma_{\rm Theory}$. In the direct comparison of Fig.~\ref{Fig V scan Gamma dir comp} (a) the differences between $\Gamma_{\rm Sim}$ and $\Gamma_{\rm Theory}$ are barely visible, whereas in Fig.~\ref{fig: Gamma comp ratios} (a) a tilt can be seen in the ratios. This tilt is the same tilt as observed for $1.2 \lesssim F_0 \lesssim 2.2$ in Fig.~\ref{Fig1: DH rates}.

In the non-zero $V$ results of Figs.~\ref{Fig V scan Gamma dir comp} (b-c) and \ref{fig: Gamma comp ratios} (b-c) the direct comparisons continue to show good agreement, and the ratios continue to show the same slight tilt. The slight deviation of $\Gamma_{\rm Sim}$ from $\Gamma_{\rm Theory}$ is quite similar to that observed for $V=0$. This similarity is not surprising as the excitation mechanism is the creation of internally independent DH pairs see Sec.~\ref{Sec: Theory and Methods - The extended Hubbard model - 2V<4t0}. However, while the change in $\Gamma_{\rm Sim}$ compared to $\Gamma_{\rm Theory}$ remains quite similar across Figs.~\ref{Fig V scan Gamma dir comp} (a-c) and \ref{fig: Gamma comp ratios} (a-c) it is clear that as $V$ increases $\Gamma_{\rm Sim}$ increases relative to $\Gamma_{\rm Theory}$. This is likely a reflection of $V$ aiding in the generation of doublons. As described in Sec.~\ref{Sec: Theory and Methods - The extended Hubbard model - 2V<4t0}, adding $V$ modifies the excitation to independent DH pairs. As $\hat{H}_V$ causes DH attraction and excitons, it stands to reason that it also decreases the energy required to create a DH pair occupying neighboring sites, even if that energy remains larger than $ \Delta_{\rm Mott}$. That energy requirement decrease will aid in the production of doublons, and as Eq.~(\ref{eq: Gamma formula}) assumes no such decrease, that will result in $\Gamma_{\rm Sim}$ increasing relative to $\Gamma_{\rm Theory}$, as observed in Figs.~\ref{Fig V scan Gamma dir comp} (a-c) and \ref{fig: Gamma comp ratios} (a-c). 

In Figs.~\ref{Fig V scan Gamma dir comp} (d) and \ref{fig: Gamma comp ratios} (d), $V=3$, meaning the system has entered the regime where the Mott exciton has separated from the continuum. This causes us to introduce two sets of results. Firstly one which keeps $F_{\rm th}=F_{\rm th}^{\rm Mott}$ and another for $F_{\rm th}=F_{\rm th}^{\rm exc}$.  In both cases, we observe worse agreement between $\Gamma_{\rm Sim}$ and $\Gamma_{\rm Theory}$, when compared to the results for $V\leq 2$. In the direct comparisons this worse agreement manifests as $\Gamma_{\rm Sim}$ being consistently higher than $\Gamma_{\rm Theory}$, whether $\Gamma_{\rm Theory}$ was obtained using $F_{\rm th}^{\rm Mott}$ or $F_{\rm th}^{\rm exc}$. This can also be seen in Fig.~\ref{fig: Gamma comp ratios} (d) as the average of the ratios is $>1$. That said $\Gamma_{\rm Theory}$ increases by using $F_{\rm th}^{\rm exc}$ in place of $F_{\rm th}^{\rm Mott}$, which means $F_{\rm th}^{\rm exc}$ gives better agreement with $\Gamma_{\rm Sim}$. However, in the ratios, Fig.~\ref{fig: Gamma comp ratios} (d), instead of the consistent slight tilt, the ratios now show a faster decrease for $F_0/F_{\rm th}\leq 0.5$ after which the ratios slowly increase. As this happens for both $F_{\rm th}=F_{\rm th}^{\rm Mott}$ and $F_{\rm th}=F_{\rm th}^{\rm exc}$ it can be concluded that Eq.~\eqref{eq: Gamma formula} is starting to fail in a way that is not fixable merely by changing from $F_{\rm th}=F_{\rm th}^{\rm Mott}$ to $F_{\rm th}=F_{\rm th}^{\rm exc}$. So while using  $F_{\rm th}=F_{\rm th}^{\rm exc}$ does bring $\Gamma_{\rm Theory}$ closer to $\Gamma_{\rm Sim}$ it does not appear to improve Eq.~\eqref{eq: Gamma formula}'s ability to capture $\Gamma_{\rm Sim}$ fundamentally. The reason for this failure is the excitons that enter the system and these decrease Eq.~\eqref{eq: Gamma formula}'s applicability.

Further increasing $V$ to $4$ gives the results shown in Figs.~\ref{Fig V scan Gamma dir comp} (e) and \ref{fig: Gamma comp ratios} (e). The excitonic states further separate from the independent DH continuum.  This distances the system from the system Eq.~(\ref{eq: Gamma formula}) was derived to handle, which is demonstrated by the larger disagreements observed. There are two main takeaways from these figures. Firstly, the dependence of $\Gamma_{\rm Sim}$ on $F_0$ is markedly different from the way $\Gamma_{\rm Theory}$ depends on $F_0$. This indicates that Eq.~(\ref{eq: Gamma formula}) fundamentally fails to capture $\Gamma_{\rm Sim}$ at these $V$ values. Secondly, the change from $F_{\rm th}=F_{\rm th}^{\rm Mott}$ to $F_{\rm th}=F_{\rm th}^{\rm exc}$ does increase $\Gamma_{\rm Theory}$ which brings it closer to the $\Gamma_{\rm Sim}$ values, but does not change the fact that Eq.~\eqref{eq: Gamma formula} fails at this point. 

Finally, in Figs.~\ref{Fig V scan Gamma dir comp} (f) and \ref{fig: Gamma comp ratios} (f), where $V=5$, $\Gamma_{\rm Sim}$ and $\Gamma_{\rm Theory}$ differ massively from one another. Note that the $\Gamma_{\rm Sim}$ results for $F_{\rm th}=F_{\rm th}^{\rm exc}$ and $F_0 < 0.2$ are missing because the value becomes negative. As described in Sec.~\ref{Sec: Theory and Methods - The extended Hubbard model - 2V approx U but 2V<U}, this is not surprising as $V=U/2$ is precisely at the phase transition point where the cost of creating a doublon by extending an exciton string is $\approx 0$. This means there is very little correlation between the excitation rate and the doublon creation rate, and therefore Eq.~\eqref{eq: Gamma formula} fails. A negative $\Gamma_{\rm Sim}$ corresponds to DH pair annihilation, which is the characteristic excitation mechanism in the CDW phase, as described in Sec.~\ref{Sec: Theory and Methods - The extended Hubbard model - 2V>U}. Note that the results in Fig.~\ref{Fig V scan Gamma dir comp} (f)  still show threshold-like behavior.

To summarize, we observe a general tendency that the formula Eq.~\eqref{eq: Gamma formula} becomes less reliable in describing the evolution of the doublon counts under quantum tunneling as excitonic effects, including the formation of CDW droplets, become significant. This may be expected given that the nature of the relevant quasiparticles is strongly modified by the nonlocal interaction $V$, compared to the case of independent doublons and holons at $V=0$. In fact, the latter cease to be relevant quasiparticles as $V$ increases.

\subsection{$\Gamma_{\rm Theory}$ and energy increase rate} \label{sec:result_C}

\begin{figure*}
    \captionsetup{justification=raggedright}
 	\includegraphics[width=\linewidth]{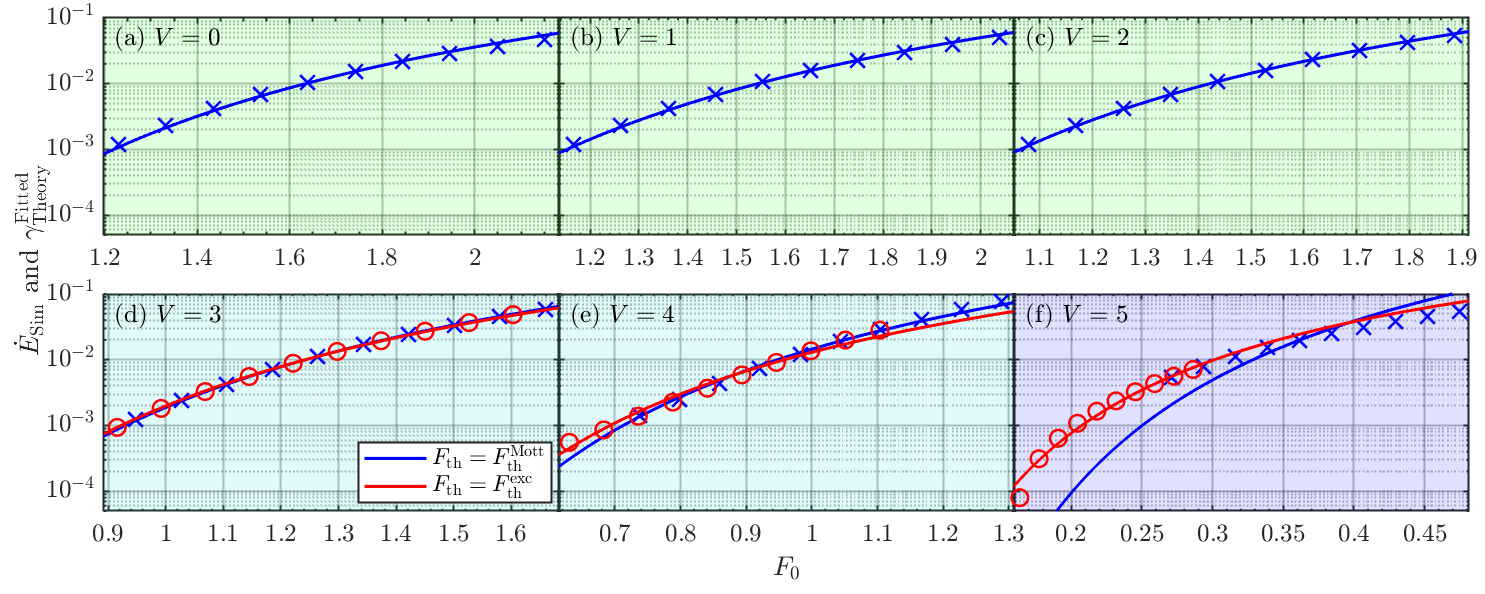}
 	\caption{Comparisons between $\dot{E}_{\rm Sim}$ and $\gamma_{\rm Theory}^{\rm Fitted}$ for all integer values of $V$ in the Mott phase and $0.4F_{\rm th}\leq F_0 \leq 0.7F_{\rm th}$. The $\dot{E}_{\rm Sim}$ values are indicated by either crosses or circles, and $\gamma_{\rm Theory}^{\rm Fitted}$ are given by the fully drawn lines. Here $\gamma_{\rm Theory}^{\rm Fitted} = C\times \Gamma_{\rm Theory}$ is $\Gamma_{\rm Theory}$ with $C$ fitted to the associated $\dot{E}_{\rm Sim}$ values. These comparisons are done using either $F_{\rm th} = F_{\rm th}^{\rm Mott}$ or $F_{\rm th} = F_{\rm th}^{\rm exc}$ as indicated in the legend. The background color indicates the regime each panel belongs to, see Fig.~\ref{Fig: schema}. $F_0$ is measured in units of $\frac{t_0}{a|e|}$, and both $\dot{E}_{\rm sim}$ and $\gamma_{\rm Theory}^{\rm Fitted}$ are in units of $\frac{t_0^2}{\hbar}$.}
 	\label{Fig V scan E dir comp}
\end{figure*}


Having now addressed when and how accurately Eq.~\eqref{eq: Gamma formula} captures the evolution of the doublon number, we move on to examine its ability to describe $\dot{E}_{\rm Sim}$, i.e., the rate of energy change in the system extracted from simulations. Extracting $\dot{E}_{\rm sim}$ is done using the same method as used to extract $\Gamma_{\rm Sim}$ from $N_{\rm D}$, i.e., we fit a straight line to the energy $E$ for $t$ between $t_r = 4$ and $t_r+1$. As the system is an infinitely long chain, we evaluate the energy per site instead of the total energy. We evaluate that energy as $E = \frac{1}{2}\left(\langle\hat{H}_i(t)\rangle + \langle\hat{H}_{i+1}(t)\rangle\right)$, where $\hat{H}_i(t)$ contains the terms of $\hat{H}(t)$ describing hops between site $i$ and $i+1$ as well as the onsite Coulomb repulsion on site $i$. We average over 2 sites to minimize the effects of the staggered magnetic field, see Sec.~\ref{Sec: Theory - methods and systems}. As discussed in Sec.~\ref{Sec: Theory and methods - The Gamma formula}, the energy evolution and the doublon number are closely related in the standard Hubbard model ($V=0$). 
We can further expect that the change in energy directly reflects the depletion of the insulating ground state due to charge carrier generation via quantum tunneling, even when the nature of the charge carriers (quasiparticles) is strongly modified from the prototypical doublons and holons. 
Namely, we expect that the functional form of Eq.~\eqref{eq: Gamma formula} remains applicable over a wide parameter range, provided that the gap size is properly adjusted.

\begin{figure}	\captionsetup{justification=raggedright}
	\includegraphics[width=\linewidth]{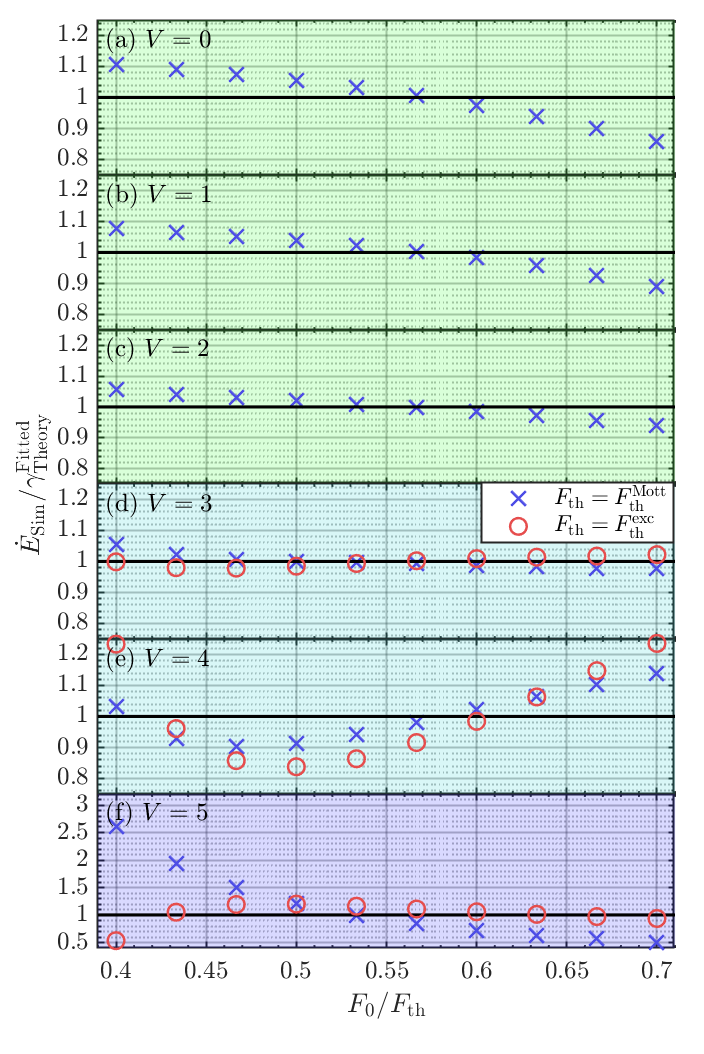}
	\caption{Ratios between $\dot{E}_{\rm Sim}$ and
$\gamma_{\rm Theory}^{\rm Fitted}$ for all integer values of $V$ in the Mott phase at $U=10$ for field strengths satisfying $0.4 \leq  F_0/F_{\rm th} \leq 0.7$ using either $F_{\rm th} = F^{\rm Mott}_{\rm th}$ or $F_{\rm th} = F^{\rm exc}_{\rm th}$ as indicated in the legend.  The
background colors indicate the regime each panel belongs to, see Fig. 1. Note the change in the ordinate in panel (f).}
	\label{Fig: V scan E}
\end{figure}

We test this conjecture using a set of figures equivalent to Fig.~\ref{Fig V scan Gamma dir comp} and \ref{fig: Gamma comp ratios}. That set is Fig.~\ref{Fig V scan E dir comp} and \ref{Fig: V scan E}. In Fig.~\ref{Fig V scan E dir comp} we show direct comparisons between $\dot{E}_{\rm Sim}$ and $\gamma_\text{Theory}^{\rm Fitted}$ and in Fig.~\ref{Fig: V scan E} we show the ratios between $\dot{E}_{\rm Sim}$ and $\gamma_\text{Theory}^{\rm Fitted}$, in both cases for all integer $V$ values in the Mott phase. Here $\gamma_{\rm Theory}^{\rm Fitted}=C\times \Gamma_{\rm Theory}$, with $C$ being a fitting parameter with units such that $\gamma_{\rm Theory}^{\rm Fitted}$ has the same units as $\dot{E}_{\rm Sim}$.  $\gamma_{\rm Theory}^{\rm Fitted}$ is fitted to either the $F_{\rm th} = F_{\rm th}^{\rm Mott}$ or the $F_{\rm th}=F_{\rm th}^{\rm exc}$, $\Gamma_{\rm Sim}$ values as indicated by the colors.  Starting with Figs.~\ref{Fig V scan E dir comp} (a) and \ref{Fig: V scan E} (a), $V=0$ shows that $\gamma_{\rm Theory}^{\rm Fitted}$ can capture $\dot{E}_{\rm Sim}$ in the regime in which Eq.~\eqref{eq: Gamma formula} was derived.
Similar agreement between $\dot{E}_{\rm Sim}$ and $\gamma_{\rm Theory}^{\rm Fitted}$ is observed in Figs.~\ref{Fig V scan E dir comp} (b) and (c) and Figs.~\ref{Fig: V scan E} (b) and (c), $V\leq 2$. This agreement indicates that $\dot{E}_{\rm Sim}$ is captured by Eq.~\eqref{eq: Gamma formula} while the system is in the dynamical regime characterized by internally independent DH pairs. This aligns with the results from comparing $\Gamma_{\rm Sim}$ to $\Gamma_{\rm Theory}$.
Increasing $V$ further to $3$ or $4$ in Figs.~\ref{Fig V scan E dir comp} (d-e) and \ref{Fig: V scan E} (d-e) does not significantly affect the degree of agreement between $\dot{E}_{\rm Sim}$ and $\gamma_{\rm Theory}^{\rm Fitted}$, contrary to the results from comparing $\Gamma_{\rm Sim}$ to $\Gamma_{\rm Theory}$. Note that in both Figs.~\ref{Fig V scan E dir comp} (d-f) and \ref{Fig: V scan E} (d-f) two sets of results are shown, one for $F_{\rm th} = F_{\rm th}^{\rm Mott}$ and another for $F_{\rm th} = F_{\rm th}^{\rm exc}$.
Contrary to what was observed when comparing $\Gamma_{\rm Sim}$ to $\Gamma_{\rm Theory}$, we observe quite good agreement between $\dot{E}_{\rm sim}$ and $\gamma_{\rm Theory}^{\rm Fitted}$ for $V > 3$. In Fig.~\ref{Fig V scan E dir comp} (d-e), $V=3$ and $4$, respectively we observe that there is very good agreement in the direct comparisons and in Fig.~\ref{Fig: V scan E} (d-e), we see that the ratios diverge slightly from one, but not nearly as much as observed in Fig.~\ref{fig: Gamma comp ratios} (d-e), particularly for $V=4$.
For $V=5$, where the CDW droplets become significant, the theoretical prediction with $F_{\rm th} = F_{\rm th}^{\rm Mott}$ shows poor agreement with $\dot{E}_{\rm Sim}$ as seen in Figs.~\ref{Fig V scan E dir comp} (f) and \ref{Fig: V scan E} (f). However, $\gamma_{\rm Theory}^{\rm Fitted}$ with $F_{\rm th} = F_{\rm th}^{\rm exc}$ shows substantial improvement and agrees well with $\dot{E}_{\rm Sim}$.


\section{Summary and outlook} \label{Sec: Summary}
In this work, we made a systematic numerical study of quantum tunneling in SCSs exposed to strong DC electric fields. 
We did this with the help of tensor-network-based methods (iTEBD and DMRG). We calculated the doublon number and the total energy in the one-dimensional extended Hubbard model, with a particular focus on the applicability range of the analytic formula Eq.~(\ref{eq: Gamma formula}) derived in Ref.\cite{Oka2012}. 
Although this formula was originally derived in the one-dimensional ($V=0$) Hubbard model, in practice it has been used beyond this limit in a wide range of theoretical~\cite{Mayer2015, Tohoyama2023, AlShafey2023, AlShafey2024} and experimental studies\cite{Mayer2015, Tohoyama2023, AlShafey2023, AlShafey2024},
hence its ability to predict both the DH pair generation rate and rate of energy increase of systems beyond this limit should be justified.
Our results can be summarized as follows.
\begin{enumerate}
\item {\bf Mott phase without exciton ($V\lesssim 2$)}:
In this case, the threshold behavior of the physical quantities (doublon number and energy) 
is observed and the analytic formula Eq.~(\ref{eq: Gamma formula}) works reasonably well.
\item {\bf Mott phase with well-defined exciton ($2\lesssim V < V_c$)}:
In this case, although the threshold behavior of the doublon number is observed, its structure tends to deviate from the analytic formula Eq.~(\ref{eq: Gamma formula}).
However, the energy rate shows notably better agreement with the formula.
If we empirically replace the Mott (band) gap with the exciton energy in $F_{\rm th}$, the prediction of the formula tends to improve, particularly in the case of the doublon number.
\item {\bf Mott phase with CDW droplet ($V \simeq U/2$)}:
Here, the energy rate shows reasonable agreement with the formula, whereas the doublon number is completely off. At this point the improvement of changing the Mott gap to the gap generated from the optical conductivity, i.e., $\Delta_{\rm exc}$ is substantial in the case of the energy rate, whereas the doublon count remains irredeemable. 
\end{enumerate}

The tendency for Eq.~\eqref{eq: Gamma formula} to become less reliable in describing the doublon number with increasing $V$ may simply arise from the fact that the nature of the charge carriers (quasiparticles) is modified, and the doublon number ceases to reflect the actual number of charge carriers. On the other hand, we have shown that the form of Eq.~\eqref{eq: Gamma formula}, with a properly chosen gap size reflecting the relevant quasiparticles, can describe the energy evolution over a wide parameter range.
These results suggest that a similar form to Eq.~\eqref{eq: Gamma formula} may be applicable to describe quantum tunneling in a broad class of one-dimensional strongly correlated insulators, although one must carefully choose the relevant physical observables. Determining the exact form of such a formula, along with a more rigorous derivation, remains an important task for future work.


Another important future tasks is to extend the numerical analysis to higher-dimensional systems and evaluate the validity of the analytic formula~\eqref{eq: Gamma formula}. In Mott insulators, carrier dynamics are highly dependent on the dimensionality of the system due to the spin-charge coupling~\cite{Murakami2023_review,Murakami2024HHG}. These numerical efforts may provide valuable insights for developing a more versatile expression of the threshold beyond the one-dimensional Hubbard limit.
Additionally, understanding tunneling behavior in SCSs beyond the paradigm of the standard Mott insulator is crucial. In Mott insulators described by the standard Hubbard model, the basic elementary excitations are doublons and holons, and the analytic formula is derived within this framework. However, as demonstrated in this paper, generic SCSs exhibit a variety of elementary excitations. Specifically, the state around $V \simeq U/2$ features CDW droplets as excited states, where the corresponding elementary excitations may represent the edges of these droplets (solitons).
Exploring phenomena beyond the standard Hubbard model paradigm would be fascinating, particularly to uncover the dynamics of emergent elementary excitations under strong fields~\cite{dziurawiec2024hhg}.

\begin{acknowledgments}
		This work was supported by the Independent Research Fund Denmark Grant No. 1026-00040B (T.H., L.B.M.) and by Grant-in-Aid for Scientific Research from JSPS, KAKENHI Grant Nos. JP21H05017 (Y.M.), JP24H00191(Y.M), and JST CREST Grant No. JPMJCR1901 (Y.M.). The iTEBD calculations have been implemented using the open-source library ITensor~\cite{ITensor_main}.
	\end{acknowledgments}

  \appendix
\section{Estimating the exciton energies with optical conductivities} \label{App: Estimating the exciton energies with optical conducitvities}
\begin{figure}
	\captionsetup{justification=raggedright}
	\includegraphics[width=\linewidth]{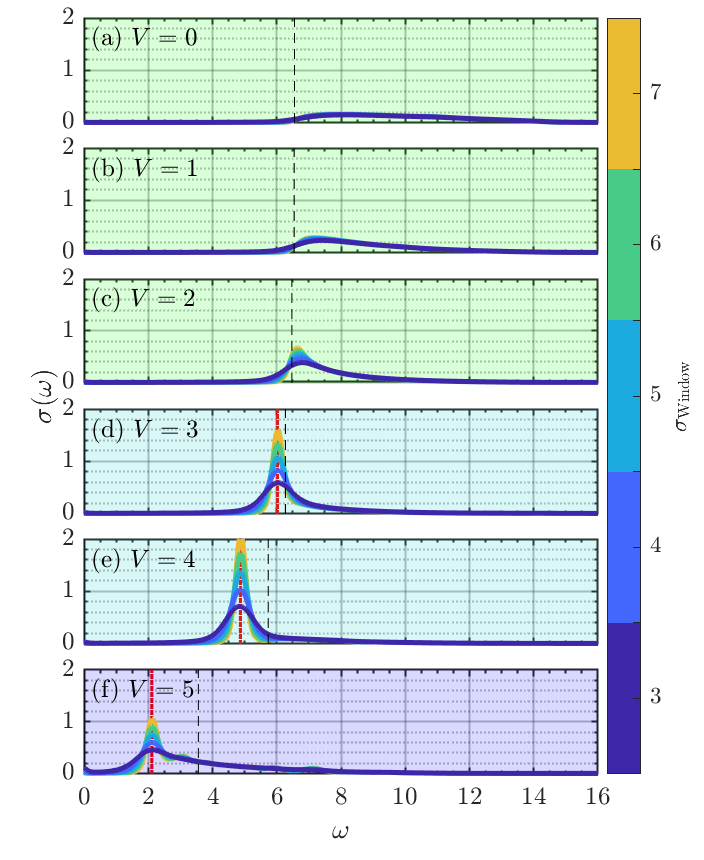}
	\caption{Real part of the optical conductivity at $U=10$ with integer values of $V$ between 0 and 5 and various values of $\sigma_\text{Window}$. The dependence on the latter shows the uncertainty on the Gaussian envelope used for the Fourier transforms in Eq.~(\ref{Eq. opt cond}). The background colors indicate the corresponding dynamical regime of the system, see Fig.~\ref{Fig: schema}. The black dashed vertical lines indicate $\Delta_\text{Mott}$ as given by Eq.~(\ref{eq: Mott gap}) and the red dotted vertical lines indicate the exciton energies. The exciton energies are only given for $V>2$ as that is where the excitonic states enter the band gap below the continuum of states with DH pairs consisting of independent doublons and holons. Here $\omega$ is given in units of $\frac{t_0}{\hbar}$, $\sigma_{\rm Window}$ is in units of $\frac{\hbar}{t_0}$, and $\sigma(\omega)$ is in units of $\frac{|e|^2 a}{\hbar}$.}
	\label{Fig: Opt cond}
\end{figure}
    
   Here we evaluate the optical conductivity to determine the exciton energies, $\Delta_{\rm exc}$, which we used to calculate $F_{\rm th}^{\rm exc}$ in the main text; Tab.~\ref{Tab: F_th exciton}.   
   The optical conductivity, $\sigma(\omega)$, can be calculated by applying a short and perturbative laser pulse to the system \cite{Murakami2023_review} as 
\begin{align}
	\sigma(\omega) &= \frac{J(\omega)}{F(\omega)} \label{Eq. opt cond},\\
	J(\omega) &= \mathcal{F}\left(\left<\hat{J}(t)\right>\right),\\
	\hat{J}(t) &= -i\sum_{i,\sigma} \left(e^{iA(t)}\hat{c}^\dagger_{i,\sigma}\hat{c}_{i+1,\sigma} - h.c.\right),
\end{align}
where $\hat{J}(t)$ is the current operator, $\mathcal{F}$ is the Fourier transform, and $F(\omega)=\mathcal{F}(F(t))$ is the Fourier transform of the electric field, $F(t)$. We note that the precise pulse shape is irrelevant to the optical conductivity given that the field strength is low enough to place the system in the linear-response regime. 
We use iTEBD to simulate the time evolution against the weak electric field. 
Because of the finite-time simulation, in practice, the Fourier transforms are done with a Gaussian envelope, $\exp(-(t-t_c)^2/(2\sigma_\text{Window}^2))$, where $t_c$ is the time of the center of the laser pulse and $\sigma_\text{Window}$ is the standard deviation on the window.

As described in Sec.~\ref{Sec: Theory}, the extended Hubbard model contains three dynamical regimes in the Mott phase at $U=10$.
These regimes are reflected in $\sigma(\omega)$. In Fig.~\ref{Fig: Opt cond} the real part of the optical conductivity is shown for integer values of $V$ between 0 and 5 and various values of $\sigma_\text{Window}$ with the dynamical regime being indicated by the background color, see Fig.\ref{Fig: schema}. The vertical dashed lines indicate the $\Delta_\text{Mott}$ values obtained from DMRG and Eq.~(\ref{eq: Mott gap}).
The red dotted lines indicate the energies of the excitonic states, $\Delta_{\rm exc}$, evaluated from the peak positions in $\sigma(\omega)$. In Fig.~\ref{Fig: Opt cond} (a-c), $V\leq 2$, we see that $\sigma(\omega)$ is largely independent of $\sigma_\text{Window}$. The main effect of increasing $\sigma_\text{Window}$ is to reduce the widening of the peak induced by the window function. Hence, if the optical conductivity does not show peaks but instead a continuum, the optical conductivity should be largely independent of $\sigma_\text{Window}$. That fits with this regime being characterized by the DH continuum resulting from the doublons and holons being independent from one another. 
We note that $\sigma(\omega)$ (a-c), $V\leq 2$, show increasingly sharp peak-like structures at the left side of the continuum.  These peak-like structures are what become clear peaks below $\Delta_\text{Mott}$ in Fig.~\ref{Fig: Opt cond} (d-e) for $4 \gtrsim V \gtrsim 2$, which correspond to Mott excitons.
Note also how that peak depends on $\sigma_\text{Window}$ and shows the expected decreasing width of the peak with increasing $\sigma_\text{Window}$. In Fig.~\ref{Fig: Opt cond} (e), $V=4$, the peak has become so tall that the continuum has become difficult to see. However, the continuum is there starting from $\omega \simeq 6$  and ending at $\omega \simeq 10$. In this case, the exciton peak is entirely distinct from the continuum due to the high binding energy of the exciton resulting from the high $V$ value. This fits with the description of this regime in the main text, Sec.~\ref{Sec: Theory and Methods - The extended Hubbard model - U>2V>4}.
At $V=5$, the system is at the transition point between the Mott and CDW phases ($V\simeq U/2$), and it enters the dynamical regime characterized by formation the CDW droplets as its excited states. Reflecting this, the optical conductivity changes dramatically as seen in Fig.~\ref{Fig: Opt cond} (f). Namely, a clear single peak below the Mott gap $\Delta_{\rm Mott}$ vanishes now, and the continuum spectrum appears there. Although the concept of a single independent exciton is not well defined in this regime, we set the first peak in $\sigma(\omega)$ as $\Delta_{\rm exc}$.

	\bibliography{this_bib_file}
	
\end{document}